\documentclass[pra,twocolumn,superscriptaddress]{revtex4}
\usepackage{graphicx}
\usepackage{epstopdf}
\usepackage{amssymb}
\usepackage{amsfonts}
\usepackage{amsmath}
\usepackage{mathrsfs}
\usepackage{bm}
\usepackage{float}
\usepackage{color}

\begin{document}

\title{Magnetization textures in twisted bilayer 2D CrX$_3$ (X=Br, I)}
\author{Feiping Xiao}
\affiliation{School of Physics and Electronics, Hunan University, Changsha 410082, China}
\author{Keqiu Chen}
\affiliation{School of Physics and Electronics, Hunan University, Changsha 410082, China}
\author{Qingjun Tong}
\email{ tongqj@hnu.edu.cn}\affiliation{School of Physics and Electronics, Hunan University, Changsha 410082, China}

\begin{abstract}
Motivated by the recent experiment demonstration of stacking dependent interlayer magnetic interaction [T. Song et al., Nat. Mater. 18, 1298 (2019); T. Li et al., Nat. Mater. 18, 1303 (2019); W. Chen et al., Science 366, 983 (2019)], we investigate the magnetization textures and the control possibilities in the moir\'{e} pattern formed of twisted bilayer two-dimensional (2D) magnets CrX$_3$ (X=Br, I). We find that the stacking dependent interlayer magnetic interaction results in the formation of periodic magnetization domains in a long-period moir\'{e} pattern. Magnetization textures with various topological numbers can be constructed, depending on the winding of the textures around the domain walls. A uniform external magnetic field competes with the lateral modulated interlayer magnetic interaction and can be utilized to tune the magnetization textures.
\end{abstract}

\maketitle

\section{Introduction}
The recent discovery of intrinsic ferromagnetism in two-dimensional (2D) van der Waals (vdW) materials provides an atomically thin arena for studying fundamental physics in 2D magnetism and building low-power spintronic devices \cite{Gong2017,Huang2017,Burch2018,Gibertini2019,Gong2019}. In layered 2D materials, the interlayer coupling is weak, which makes it possible to manipulate the interlayer magnetic order in layered 2D magnets by electric gating or magnetic fields \cite{JiangNM2018,Huang2018,WangNN2018,Deng2018}. This tunability enables the fabrication of atomically thin magnetic tunnel junction devices, with giant tunneling magnetoresistance values being measured experimentally \cite{Song2018,Klein2018,WangNC2018,Ghazaryan2018,Kim2018,SongNL2019}. When stacked with other nonmagnetic layered materials, magnetic order can be imprinted in the nonmagnetic material via the magnetic proximity effect, which provides an efficient platform for spin control and quantum state engineering \cite{Qiao2014,Qi2015,Zhang2018,?uti?2019,Zhao2017,Scharf2017,Zhong2017,Seyler2018,Zhong2020,Ciorciaro2020}. For instance, the spin and valley pseudospin in monolayer transition metal dichalcogenides have been demonstrated to be controlled by stacking on chromium triiodide (CrI$_3$) \cite{Zhong2017,Seyler2018,Zhong2020} and chromium tribromide (CrBr$_3$) \cite{Ciorciaro2020}. Recently, topological spin texture has been reported in vdW layered magnet Fe$_3$GeTe$_2$ \cite{TPark2019,YWu2019,HWang2019,BDing2020,MYang2020} and Cr$_2$Ge$_2$Te$_6$ \cite{MGHan2019}, and also predicted theoretically in monolayer CrI$_3$ \cite{JLiu2018,AKBehera2019,XLu2020} and the Janus magnet \cite{JYuan2020}.

The physical properties of vdW layer materials depend sensitively on the interlayer interaction, which is further connected to the interlayer stacking order of the atomic structure. For example, in bilayer graphene, the low-energy state in AA configuration features linear dispersion, while the one in AB configuration is parabolic \cite{Park2015}. In bilayer transition mental dichalcogenides, both resonance Raman and photoluminescence spectra show distinct features in 3R and 2H stacking configurations \cite{Xia2015}. The stacking dependence is more prominent in vdW 2D magnets, in which the stacking order is directly related to the arrangement of magnetic momentums of the two layers. Indeed, in bilayer CrI$_3$, first-principles studies have revealed a rich phase diagram between interlayer ferromagnetic (FM) and antiferromagnetic (AFM) orders as the stacking configuration is continuously changed \cite{WangNC2018,Sivadas2018,JiangPRB2019,Soriano2019,JangPRM2019}. This interesting stacking dependent interlayer magnetism has been confirmed experimentally in CrI$_3$ \cite{SongNM2019,Li2019}, and also demonstrated in CrBr$_3$ \cite{Chen2019}, suggesting a way for designing magnetism via controlling stacking order. The stacking dependent interlayer magnetic coupling is expected to be general in magnetic vdW materials, as the magnetic interaction depends intimately on both relative distance and orientation between the magnetic moments.

Moir\'{e} pattern ubiquitously forms in vdW layered materials due to the misorientation or lattice mismatch between the constituent layers \cite{Geim2013,Yankowitz2012,Zhang2017,Jung2014,Tong2017,Yu2017,Wu2018,Seyler2019,Tran2019,Jin2019,Alexeev2019}. In a long-period moir\'{e} pattern, the interlayer atomic configuration resembles lattice commensurate structure, while the stacking order changes smoothly over long range. Because the interlayer interaction generally depends on the stacking order, such a spatial change of the stacking order then introduces a lateral modulation of interlayer interaction in the moir\'{e} pattern. It has been suggested that such a spatially changing interlayer magnetic interaction in 2D magnets can be used to generate various magnetization textures  \cite{Tong2018,Hejazi2020,Hejazia2020,MAkram12020}. Extending these ideas to realistic material system would facilitate their experimental observation and potential application in device fabrication.

In this work, we show that, nonuniform magnetization textures can be formed in the moir\'{e} pattern of twisted bilayer 2D ferromagnet CrI$_3$ and CrBr$_3$ arising from the stacking dependent interlayer magnetic interaction. In a moir\'{e} unit cell, there are three magnetization domains surrounded by the magnetization textures of opposite direction. These magnetization textures can be engineered to have topological number ranging from 0 to 3. An external magnetic field competes with the stacking dependent interlayer magnetic interaction and can switch the three magnetization domains to the ones of opposite direction, creating a symmetric hysteresis with double loops in the magnetic dynamics. Arising from the topological protection, the topological number is conserved during the whole process. These interesting magnetization textures can be generated via annealing from a paramagnetic state. Our results suggest new possibilities for the investigation of 2D magnetism and suggest a route towards nanoscale magnetization textures by moir\'{e} engineering.

The rest of the paper is organized as follows. In Sec. II, we give a brief account of atomic structure of the moir\'{e} pattern in bilayer ferromagnets. The stacking dependent interlayer magnetic interaction is also given by first-principles calculations. Different magnetization textures and their topological properties are present in Sec. III. The magnetic field dependence of the magnetization textures and the dynamics are given in Sec. IV. Finally, a discussion on the preparation of the magnetization textures and a summary are given in Sec. V.

\begin{figure}[tbp]
\centering
\includegraphics[width = 1.0\columnwidth] {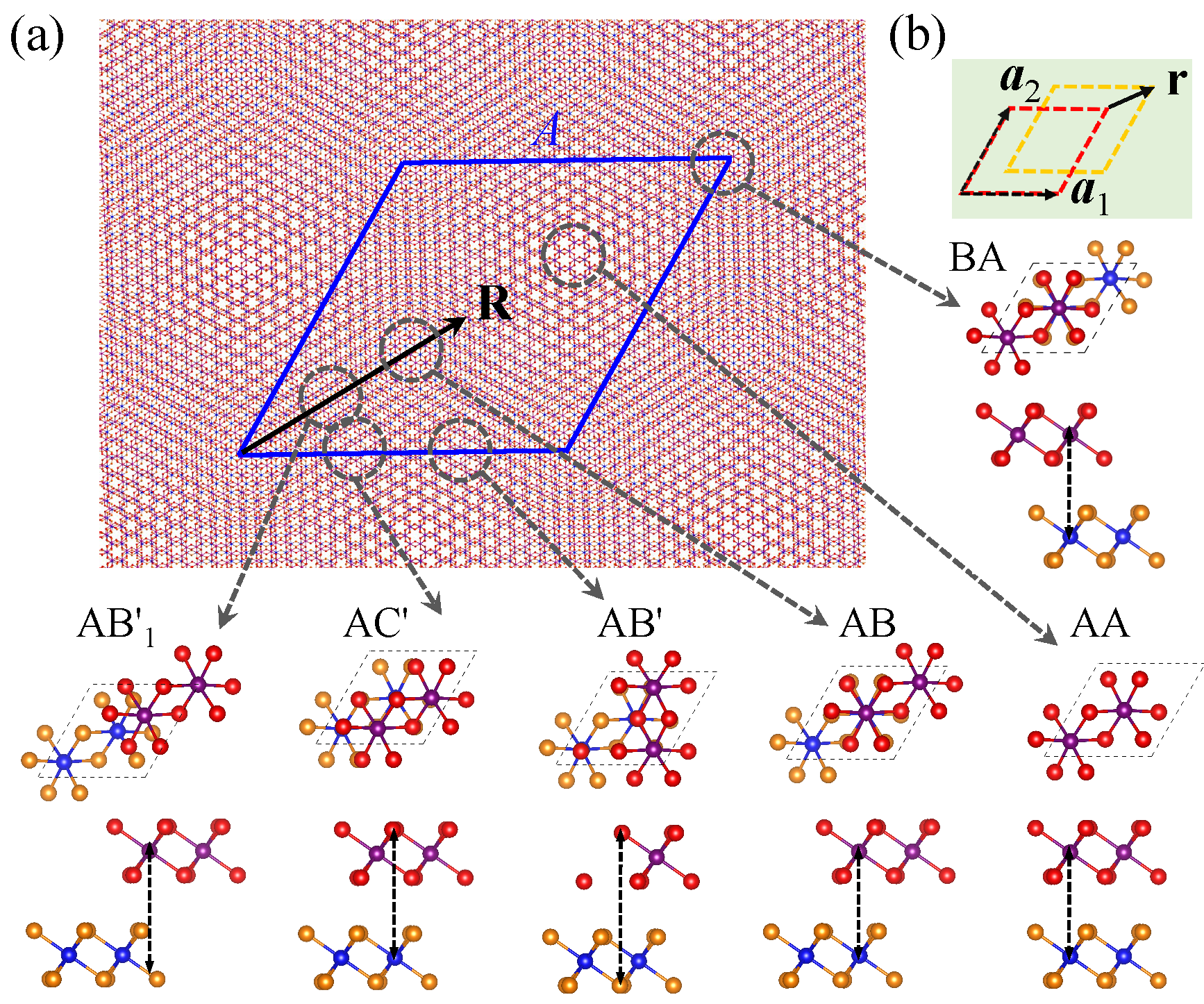}
\caption{(a) Moir\'{e} pattern of bilayer CrX$_3$ (X=Br, I) with a twisting angle of $3^{\circ}$. The blue rhombus represents a moir\'{e} unit cell with periodicity \emph{A}. Six local regions resembling lattice commensurate bilayers are named as AB$^{'}_1$, AC$^{'}$, AB$^{'}$, AB, AA, and BA stacking respectively. At each local, the top and side views are given. The red (orange) and purple (blue) balls represent X and Cr atoms of top (bottom) layer respectively. (b) The interlayer atomic registry at local region \textbf{R} in the moir\'{e} pattern is characterized by an interlayer translation vector \textbf{r} between two layers. } \label{fig1}
\end{figure}

\section{atomic structure of the moir\'{e} pattern and stacking dependent interlayer magnetic interaction}\label{model}
Monolayer CrX$_3$ (X=Br, I) has a honeycomb lattice structure possessing a magnetic moment of $1.5\mu _{B}$  per Cr atom \cite{Zhang2015}. For bilayer CrX$_3$, a small twisting or strain between the layers creates a long-period moir\'{e} pattern, as schematically shown in the Fig. 1. The moir\'{e} periodicity is approximately $A\approx a/\sqrt{\delta ^{2}+\theta ^{2}}$ for small lattice mismatch $\delta$  and/or twisting angle $\theta$, where $a$ is the lattice constant of the monolayer. Because the periodicity $A$ can be further tuned by a relative twisting or strain between the layers, it is taken as a variable in the following. In a long-period moir\'{e} pattern, the stacking order in each local region \textbf{R} is similar to the lattice-matched stacking configuration but changes smoothly over long range. Typical bilayer configurations are shown in Fig. 1, which are named as AB$^{'}_1$, AC$^{'}$, AB$^{'}$, AB, AA, and BA stacking in the following. The AB and BA stackings are the same as rhombohedral stacking, and the AB$^{'}$-stacking is the same as monoclinic stacking \cite{McGuire2015}. In Ref. \cite{Chen2019}, a point close to AB$^{'}_1$ stacking in CrBr$_3$ is named as R-type stacking. The interlayer stacking order at any local region \textbf{R} in the moir\'{e} pattern can be characterized by an interlayer translation vector \textbf{r}, which is defined in a monolayer unit cell (c.f. Fig. 1(b)). The stacking configurations of AB$^{'}_1$, AC$^{'}$, AB$^{'}$, AB and AA correspond to the top layer laterally shifted by $\textbf{r}= \left \{ \frac{\bm{a}_{1}+\bm{a}_{2}}{6},\frac{\bm{a}_{1}}{3},\frac{2\bm{a}_{1}}{3},\frac{\bm{a}_{1}+\bm{a}_{2}}{3},\frac{2(\bm{a}_{1}+\bm{a}_{2})}{3} \right \}$ with respect to bottom layer originated from BA configuration, where $\bm{a}_{1}$ and $\bm{a}_{2}$ are the two unit vectors of the monolayer.

In a moir\'{e} pattern, the number of atoms involved is $\propto \left ( A/a \right )^{2}$, therefore a direct calculation of its magnetic property is challenging. However, because the atomic registry in the moir\'{e} pattern changes smoothly, each local region resembles lattice matched commensurate structure. Based on this observation, in the following we first study the interlayer magnetic interaction at each local region and then extend to the whole moir\'{e} pattern by introducing an effective magnetic field.

\begin{figure}[tbp]
\centering
\includegraphics[width = 1.0\columnwidth] {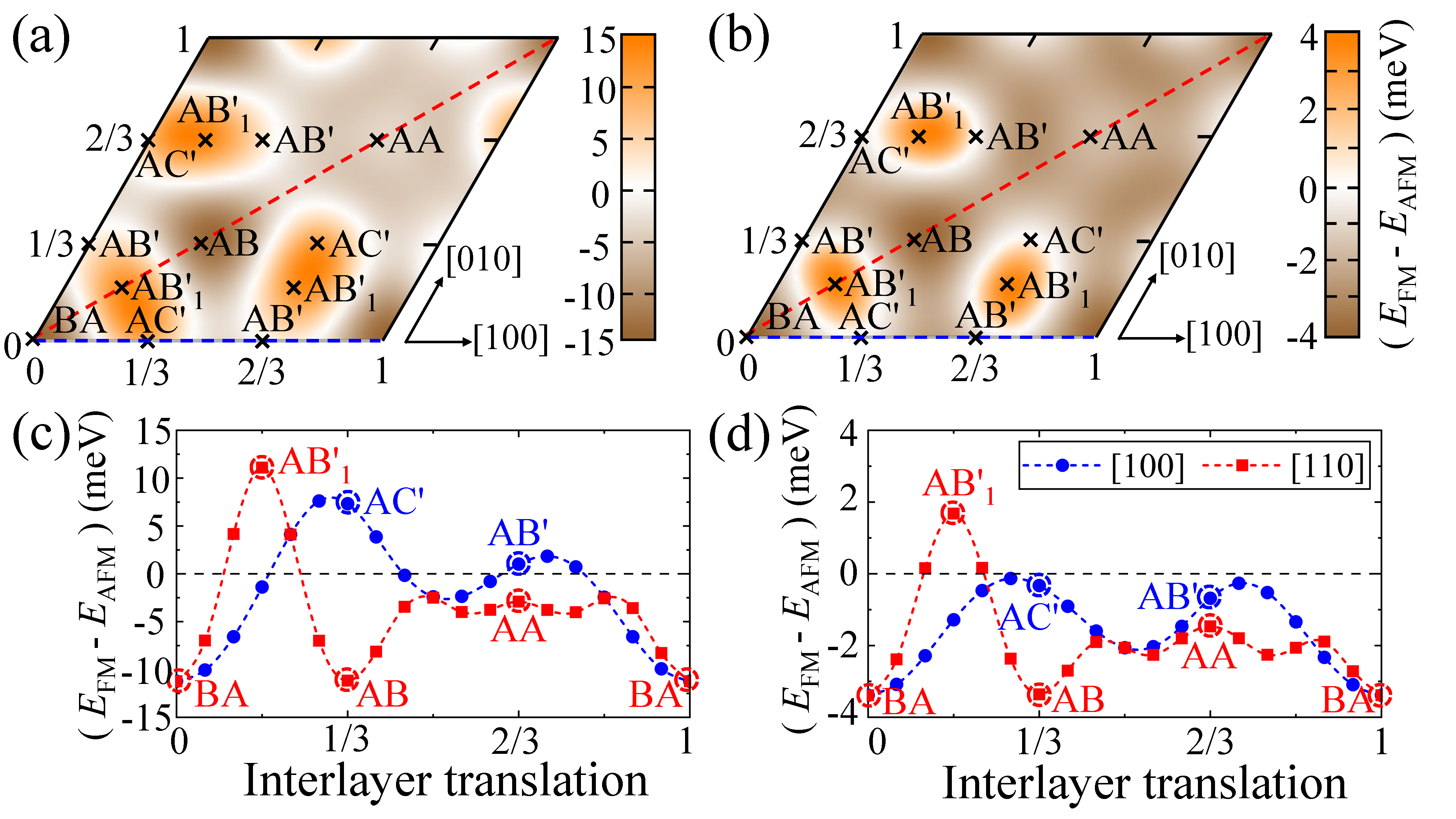}
\caption{ Energy difference between interlayer FM and AFM states of bilayer CrI$_3$ (a) and CrBr$_3$ (b) as a function of interlayer translation in a unit cell. The color-maps are obtained by interpolating the data points on a $12\times 12$ grid. The stacking configurations listed in Fig. 1 are also marked. (c, d) Detailed variation of energy difference between interlayer FM and AFM states as a function of interlayer translation along [100] (blue-dashed line with dots) and [110] (red-dashed line with squares) direction for CrI$_3$ and CrBr$_3$. } \label{fig2}
\end{figure}

The interlayer stacking-order dependent magnetic interaction is related to energy difference between interlayer FM and AFM spin configurations, which can be obtained from first-principles calculations and are shown in Figs. 2(a) and 2(b) for CrI$_3$ and CrBr$_3$ respectively. The details of the first-principles calculations and structure parameters are given in Appendix A. One can see that the ground state modulates between FM and AFM as the interlayer stacking order changes. For both CrI$_3$ and CrBr$_3$, the three \emph{C}$_3$-rotation symmetric stackings (AA, AB and BA) favor FM state. However, the magnetic interlayer interaction for \emph{C}$_3$-rotation symmetry broken configurations are different for these two materials. For CrI$_3$, the AB$^{'}_1$, AB$^{'}$ and AC$^{'}$ regions favor AFM ground state, while for CrBr$_3$, the AB$^{'}_1$ region favor AFM ground state, however AB$^{'}$ and AC$^{'}$ regions favor FM ground state. These results are consistent well with the recent experiment observation of stacking-dependent interlayer magnetism, where the monoclinic (AB$^{'}$) stacking in CrI$_3$ \cite{SongNM2019,Li2019} and R-type (AB$^{'}_1$) stacking in CrBr$_3$ \cite{Chen2019} has been confirmed to possess AFM ground state. Figs. 2(c) and 2(d) give a detailed variation of the energy difference between interlayer FM and AFM states along the [100] (blue-dashed line with dots) and [110] (red-dashed line with squares) direction of a unit cell, which includes all the above six stacking configurations. One can see that the magnitude of  interlayer magnetic interaction of CrI$_3$ is stronger than that of CrBr$_3$. Such a stacking order dependent interlayer magnetic interaction, combined with the dramatic tunability of the former by a mechanical means, point to exciting possibilities to tailor the magnetic states in the atomically thin vdW 2D magnets \cite{SongNM2019,Li2019,Chen2019}.

\begin{figure*}[tbp]
\centering
\includegraphics[width = 1.6\columnwidth] {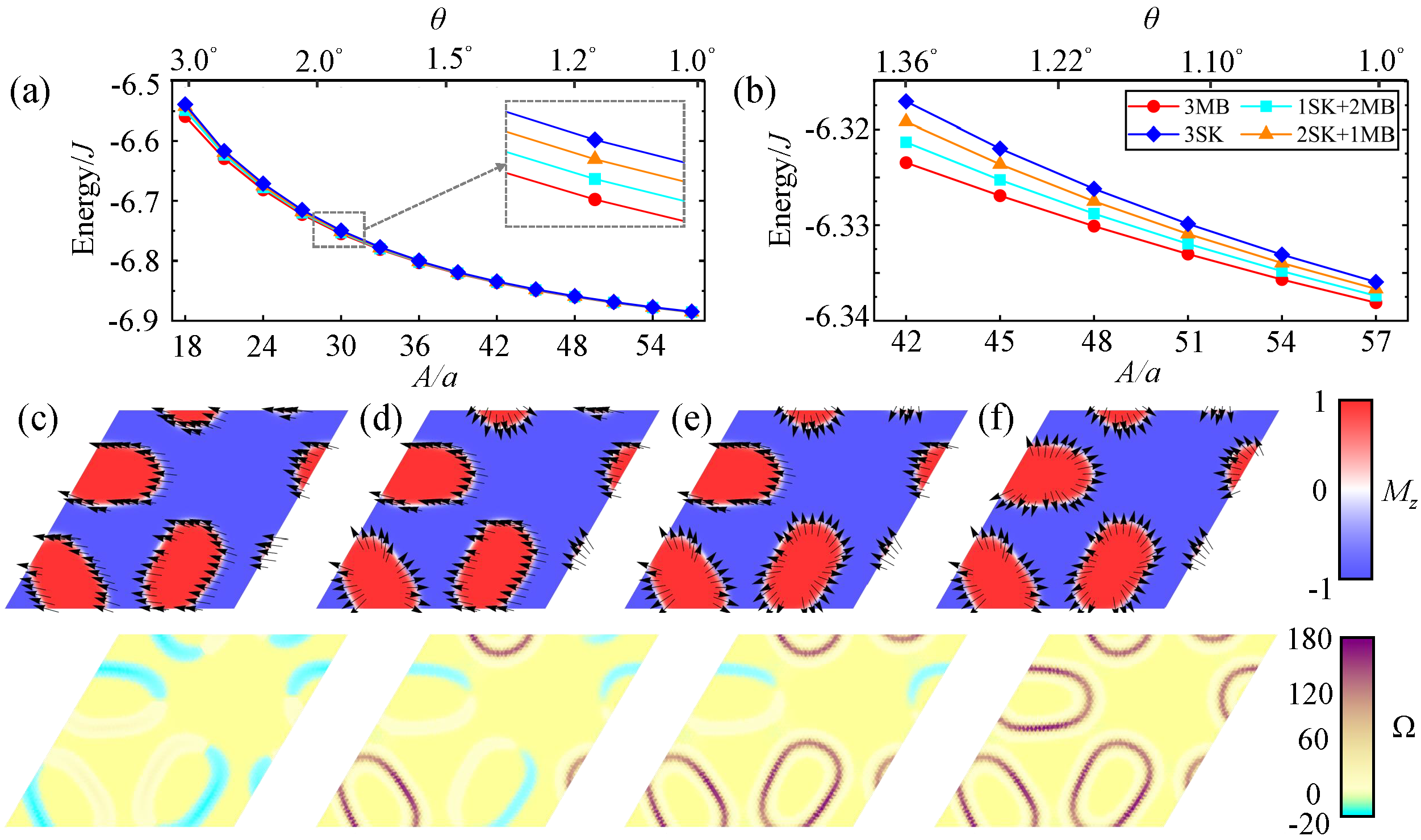}
\caption{ Energy as a function of moir\'{e} periodicity for twisted bilayer CrI$_3$ (a) and CrBr$_3$ (b) with magnetization textures of 0 SK (3MB), 1 SK (1SK+2MB), 2 SK (2SK+1MB) and 3 SK (3SK) in a moir\'{e} supercell. The corresponding magnetization textures with SK number of 0-3 in a moir\'{e} unit cell are shown in (c-f). The color-map represents the out-of-plane magnetization and the arrows show the in-plane one. The low panels show the distribution of the emergent electromagnetic field $\Omega$, whose integration over a moir\'{e} unit gives topological charge 0-3 for (c-f) respectively. } \label{fig3}
\end{figure*}

\section{magnetization textures in the moir\'{e} pattern}\label{etd}
With the knowledge of stacking dependence of interlayer magnetic interaction in the commensurate bilayers, we now turn to study the possible magnetization textures in the moir\'{e} pattern of 2D magnets. We assume that the magnetic order in one of the two layers is fixed, for example by a substrate. We name the two layers as fixed and free layers in the following. Then at different locals the magnetic order in the free layer tends to point at different directions depending on the interlayer magnetic interaction. Explicitly, if the interlayer interaction is FM, the magnetic order in the free layer tends to align parallel with the one in the fixed layer. On the other hand, if the interlayer interaction is AFM, it tends to align antiparallel with the one in the fixed layer. The effect of a fixed layer on the free layer then can be regarded as an effective magnetic field $\textbf{B}(\textbf{r})$, which is stacking order
\textbf{r} dependent. This effective field can be calculated from
\begin{equation}
E_{\textup{FM}}-E_{\textup{AFM}}=-4\textbf{B}(\textbf{r})\cdot \textbf{m}
\end{equation}
where $E_{\textup{FM}}$ and $E_{\textup{AFM}}$ are the energies of the interlayer FM and AFM states in a unit cell and $\textbf{m}=1.5\mu _{B}\textbf{z}$ is magnetic moment of Cr atom. Note that the effective field is stacking dependent, which aligns with the magnetization of the fixed layer when the ground state favors FM order, while anti-aligns when the ground state favors AFM order.

In a moir\'{e} pattern, the interlayer stacking order changes smoothly, and one expects that the spatially modulated interlayer magnetic interaction (c.f. Fig. 2) then defines a spatially modulated effective magnetic field for the free layer. We assume that the moir\'{e} pattern is formed by two rigid lattices (i.e. no lattice reconstruction). In this case, the mapping between the local registry (characterized by the interlayer translation vector $\textbf{r}$ defined in Fig. 1b) and the location $\textbf{R}$ in the moir\'{e} pattern is a linear transformation \cite{Jung2014,Stephen2018}. For small twisting angle $\theta$ and biaxial strain $\delta$, the mapping functions take the form of $\textbf{r}\left ( \textbf{R} \right )=\begin{pmatrix}
 0&-\theta  \\
 \theta & 0
\end{pmatrix}\textbf{R}$ and $\textbf{r}\left ( \textbf{R} \right )=\begin{pmatrix}
 \delta &0  \\
 0 & \delta
\end{pmatrix}\textbf{R}$ respectively. We have checked that these two types of mapping functions give almost the same results. In the following, we use the mapping form of biaxial strain to define the moir\'{e} magnetic field $\textbf{B}(\textbf{R}(\textbf{r}))$. Under this local approximation, the magnetization of free layer can be simulated with the effective Hamiltonian including the spatially changing effective magnetic field \textbf{B}(\textbf{R}$_{i}$) and intralayer magnetic interactions,
\begin{equation}
\begin{split}
H=-J\sum_{<i,j>}\textbf{m}_{i}\textbf{m}_{j}-K\sum_{i}(\textup{m}_{z,i})^{2}
\\-\sum_{i}[\textbf{B}(\textbf{R}_{i})+\textbf{B}_{\textup{ext}}]\cdot\textbf{m}_{i}
\end{split}
\end{equation}
where $\left \langle i,j \right \rangle$ covers all nearest neighboring sites of the hexagonal lattice and $\textbf{m}_{i}$ is magnetic moment at \emph{i}-th Cr site. \emph{J} is the intralayer exchange coupling and \emph{K} is the magnetic anisotropic energy, which are given in the Table II in Appendix A. We have also added a uniform external magnetic field  $\textbf{B}_{\textup{ext}}$ to tune the magnetic configuration. The dipolar interaction is orders weaker than the exchange interaction and neglected here. Via introducing this moir\'{e} magnetic field $\textbf{B}(\textbf{R})$, the bilayer moir\'{e} system is simplified to a monolayer honeycomb lattice experiencing a spatially changing external magnetic field of moir\'{e} periodicity $A$, which is discretized on the monolayer honeycomb lattice in moir\'{e} scale.

The spatially modulated effective magnetic field tends to create nonuniform magnetization domains. While the intralayer exchange and anisotropic interaction favor uniform magnetization texture. Their competition determines the final magnetization configuration. The steady-state and dynamics of the magnetization textures can be solved from the Landau-Lifshitz-Gilbert equation,
\begin{equation}
\frac{\mathrm{d} \textbf{m}_{i}}{\mathrm{d} x}=-\gamma \textbf{m}_{i}\times \textbf{H}_{i}^{eff}+\alpha \textbf{m}_{i}\times \frac{\mathrm{d} \textbf{m}_{i}}{\mathrm{d} x}
\end{equation}
where $\textbf{H}_{i}^{eff}=-\frac{\partial H}{\partial \textbf{m}_{i}}$, and $\gamma$ and $\alpha$ are the gyromagnetic ratio and Gilbert damping coefficient respectively. We obtain various magnetization textures by relaxing from different initial magnetization configurations. The initial magnetization configurations with different topological numbers can be designed via shaping the in-plane magnetic moments around the domain walls to align parallel or form a vortex. $\alpha =1$ is used for faster convergence and the periodic boundary condition is also used in all of the following calculations.

When atomic lattice relaxation is considered, the mapping between the local atomic registry and moir\'{e} superlattice would become nonlinear. Furthermore, when the in-plane vector in local registry space becomes divergent, the atomic configuration in moir\'{e} space would have rotation \cite{Stephen2018,Shiang2019}. We also note that the moir\'{e} pattern is quasi-periodic, in which the atomic structures are slightly different in different moir\'{e} unit cells. These effects would change the distribution of effective magnetic field $\textbf{B}(\textbf{r})$ in the moir\'{e} pattern. The lattice reconstruction would also change the relative positions of neighboring magnetic moments used for the numerical calculations, which are fixed for the moir\'{e} pattern formed by two rigid lattices. Including these lattice reconstruction effects should take into account the competition between the adhesion and the elastic properties of the 2D magnets, which would be an interesting topic for future studies. When the twisting angle is large, the atomic reconstruction is weak and the linear mapping is valid \cite{Stephen2018,Nam2017}.

In the moir\'{e} of bilayer magnets, we indeed find the formation of nonuniform magnetization textures, including topologically trivial magnetic bubble (MB) and magnetic skyrmion (SK) with different topological numbers, as shown in Figs. 3(c-f). We name the magnetization aligned or anti-aligned with the one in fixed layer as aligned or anti-aligned magnetization in the following. One can see that there are three anti-aligned (red) domains, centered at the AB$^{'}$$_1$ stacking regions. For CrBr$_3$, the uniform FM state is unstable and energetically higher than the MB state for a moir\'{e} periodicity larger than 42\emph{a}, corresponding to a twisting angle of  $\sim 1.4^{\circ}$. For CrI$_3$, MB texture forms for a smaller moir\'{e} with a periodicity of 18\emph{a}, arising from the much larger interlayer magnetic interaction (c.f. Fig. 2). Figs. 3(a) and 3(b) show that, as the periodicity increases, the energies of all of the four nonuniform textures decrease. This is because in a long-period moir\'{e}, the magnetization orders smoothly that reducing the intralayer exchange energy and is hence dominant by the interlayer moir\'{e} magnetic interaction, which has a sign change in the moir\'{e} supercell. Therefore, the periodic magnetization domains are more stable in a long-period moir\'{e}.

\begin{figure*}[tbp]
\centering
\includegraphics[width = 1.4\columnwidth] {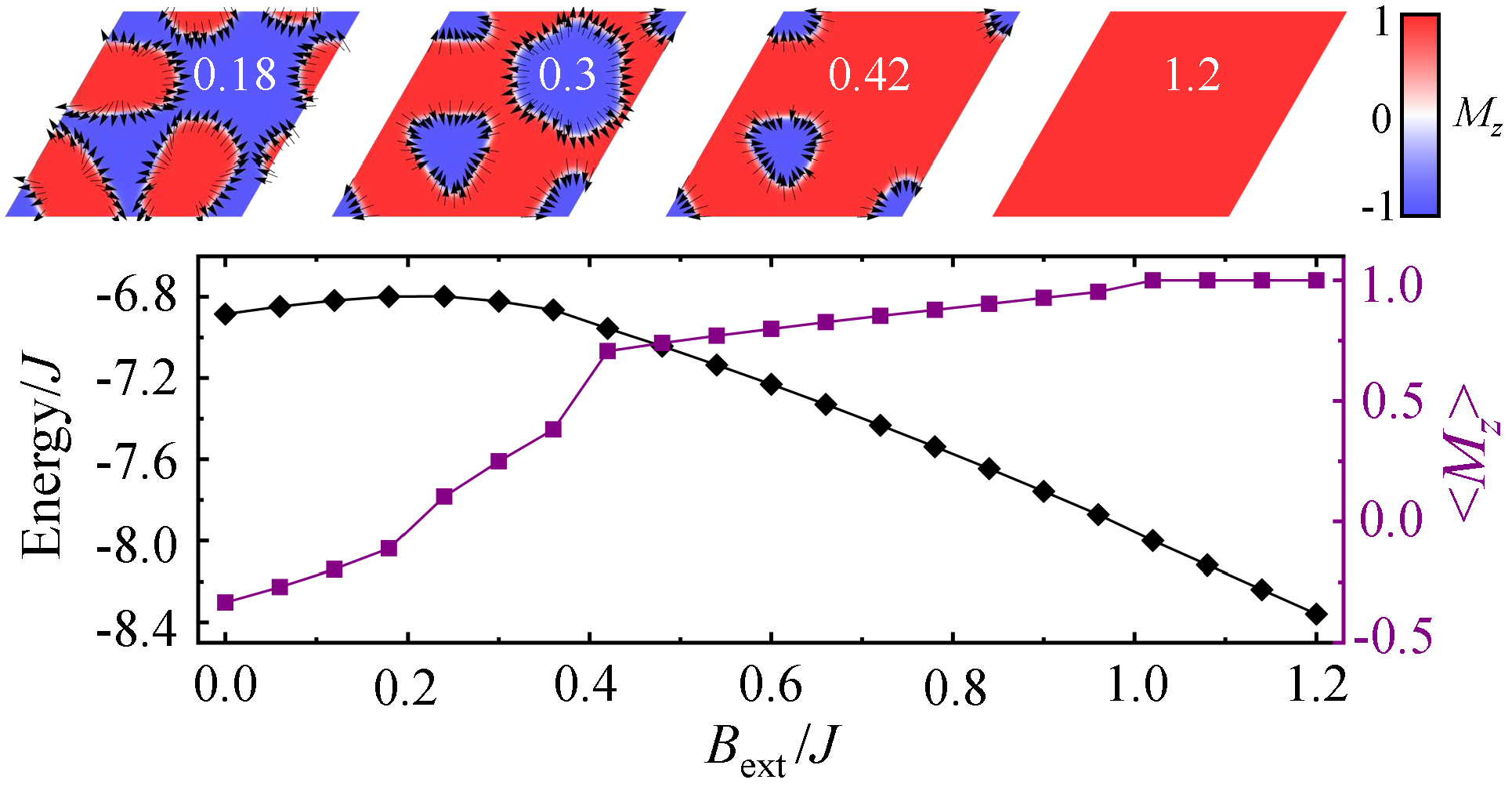}
\caption{ Energy (black line with diamonds) and averaged out-of-plane magnetization $\left \langle M_{z} \right \rangle$ (purple line with squares) for 3SK magnetization texture of twisted bilayer CrI$_3$ in a moir\'{e} supercell with  $A=57a$ under an external magnetic field along \emph{z}-direction. Four magnetization textures of $B_{\textup{ext}}/J=\left \{ 0.18,0.3,0.42,1.02 \right \}$  are shown on the top panel.  } \label{fig4}
\end{figure*}

Figs. 3(c-f) present four typical low-energy magnetization textures, which possess three anti-aligned magnetization domains surrounded by the aligned magnetization texture. Around the domain walls, the magnetization textures have different winding manner, i.e. either aligning parallel or forming a vortex. We note that the formation of nonuniform in-plane magnetization in SK state costs more intralayer exchange energy, leading to a higher energy than MB state and the energy increases linearly with the increase of SK number, as shown in Figs. 3(a) and 3(b). Therefore the topologically trivial MB state is always the lowest energy magnetization texture. These nontrivial magnetization textures can generate an emergent electromagnetic field defined as \cite{Nagaosa2013}
\begin{equation}
\Omega \left ( \textbf{r} \right )= \frac{1}{2}\textbf{m}\cdot \left ( \frac{\partial \textbf{m}}{\partial x} \times \frac{\partial \textbf{m}}{\partial y}\right )
\end{equation}
The lower panels in Figs. 3(c-f) show the emergent electromagnetic field $\Omega\left ( \textbf{r} \right )$, which is mainly distributed around the domain walls manifesting the changing of magnetization textures. One can see that $\Omega\left ( \textbf{r} \right )$ in a domain wall with vortex structure is much larger than the one when parallelly aligned, demonstrating that the former possesses a heavier deformation and costing more intralayer exchange energy. When transporting in such a magnetization texture, carriers would experience a topological Hall effect \cite{Nagaosa2013}. The integration over the moir\'{e} unit cell gives a quantized topological charge $C=\frac{1}{2\pi }\iint\Omega\left ( \textbf{r} \right )d^{2}\textbf{r}$. Our calculation shows that the corresponding topological charges are 0, 1, 2 and 3 for configurations (c-f) respectively. We name these configurations as 3MB, 1SK+2MB, 2SK+1MB and 3SK accordingly. Note that there are other configurations of different topological charge, which are energetically higher and can be found via designing the initial configurations with higher winding number around the domain walls. These higher topological number skyrmions can also be created via annealing from a paramagnetic state with randomly aligned magnetic moments, as discussed in Sec. V.

\begin{figure*}[tbp]
\centering
\includegraphics[width = 1.4\columnwidth] {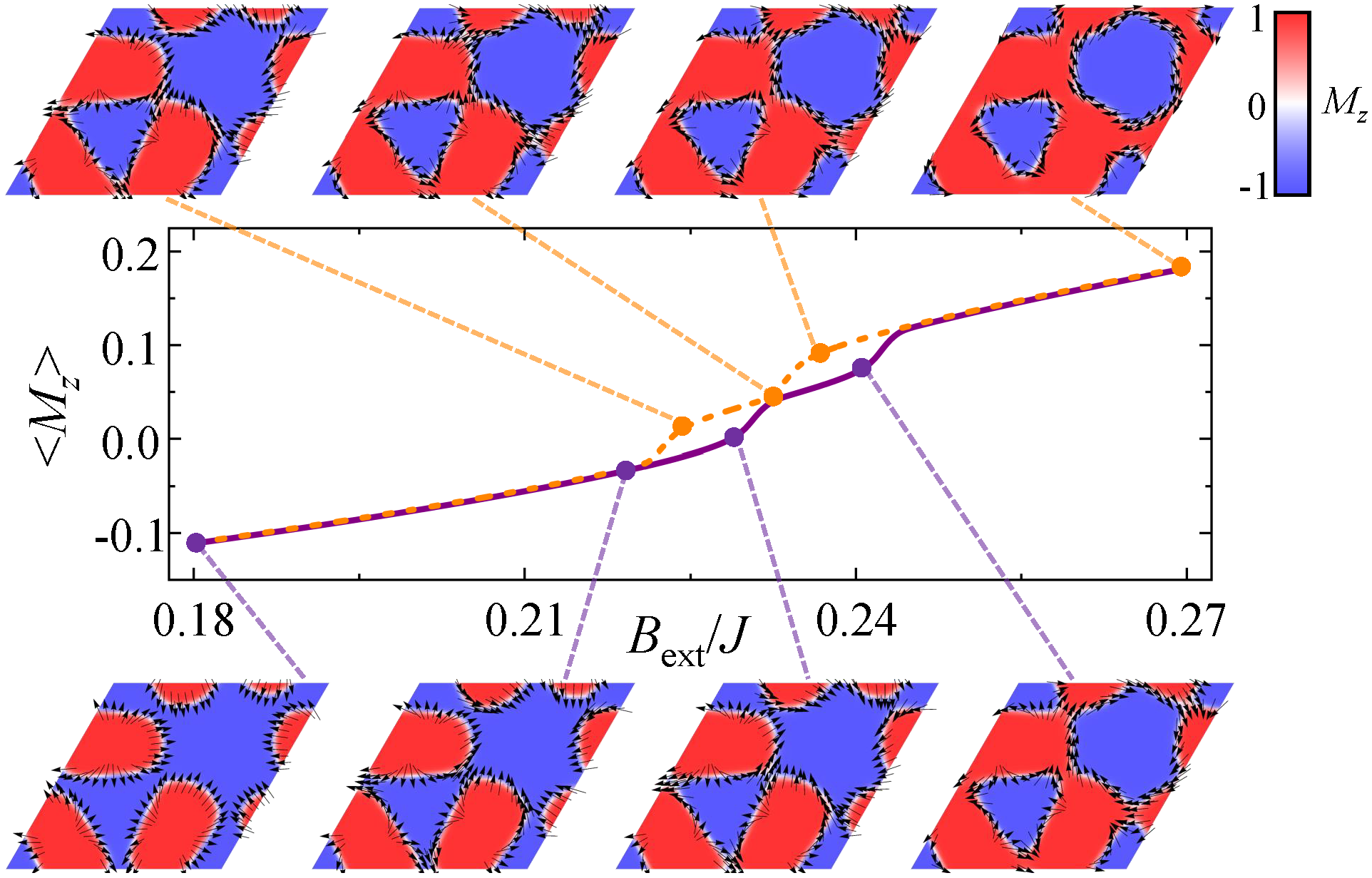}
\caption{ Magnetic hysteresis for 3SK magnetization texture of twisted bilayer CrI$_3$ in a moir\'{e} supercell with $A=57a$  when slowly increasing (purple-solid line) and then decreasing (orange-dashed line) the external magnetic field $B_{\textup{ext}}$. The magnetic configurations at representative magnetic field values are plotted.  } \label{fig5}
\end{figure*}

\section{magnetic control on the magnetization texture}\label{cd}
An external magnetic field $B_{\textup{ext}}$ along \emph{z}-direction prefers uniform spin polarized texture, which would compete with the spatially changing effective field and provide an efficient means to tune the nonuniform magnetization textures. Fig. 4 shows the steady-state magnetization structure when an external magnetic field is applied for 3SK magnetization texture (see Appendix B for magnetic control of other magnetization textures). With the increase of the magnetic field, the three anti-aligned domains first become larger and then merge into each other, with three aligned domains appearing centered at AA, AB and BA stacking regions for $B_{\textup{ext}}=0.3J$. When further increasing the magnetic field, the aligned domain around AA stacking disappears first, and then the domains at AB and BA regions disappear. This is because the interlayer FM magnetic interactions in AB and BA regions are larger than the one in AA region. For $B_{\textup{ext}}> 1.02J$, the magnetization texture becomes a spin polarized state. With the increase of the magnetic field, the energy of magnetization texture (black line with diamonds in Fig. 4) first increases. This is because the Zeeman energy $-B_{\textup{ext}}\left \langle M_{z} \right \rangle$ is positive for a negative net magnetization $\left \langle M_{z} \right \rangle$ (purple line with squares). When $\left \langle M_{z} \right \rangle$ become positive for $B_{\textup{ext}}> 0.24J$, the Zeeman energy become negative and hence the energy deceases with $B_{\textup{ext}}$ now.

To study in detail the transition from the three anti-aligned domains to the three aligned domains in the presence of an external magnetic field, we plot the magnetic dynamics when slowly sweeping up (purple-solid line) and then down (orange-dashed line) the magnetic field over a small range (from 0.18\emph{J} to 0.27\emph{J}) in Fig. 5. With an applied uniform external magnetic field $B_{\textup{ext}}$, the total effective field felt by the monolayer magnet changes to $B_{\textup{ext}}+B(\textbf{R})$, which drives slowly the magnetization dynamics when $B_{\textup{ext}}$ changes adiabatically. The intralayer exchange and anisotropic energy tend to keep the initial shape of the originally aligned magnetizations. On the other hand, the slowly changing effective magnetic field tends to drive the magnetization to follow its dynamics, which would flip the aligned magnetizations and merge the three disconnected anti-aligned domains. For small external magnetic field $B_{\textup{ext}}$, the former effect dominates and the three anti-aligned domains are still disconnected. As the magnetic field increases beyond some critical value, the aligned magnetizations at the merging areas flip and the three anti-aligned domains merge quickly with the formation of disconnected aligned domains. In the reversal process, the anti-aligned magnetizations also tend to keep their original shapes. When the magnetic field decreases beyond some critical value, the anti-aligned magnetizations at the merging areas flip and the aligned domains merge quickly with the formation of disconnected anti-aligned domains. The averaged magnetization $\left \langle M_{z} \right \rangle$ in the whole process forms a magnetic hysteresis loop. Interestingly, the averaged magnetization $\left \langle M_{z} \right \rangle$ shown in Fig. 5 forms a double hysteresis loop, with the lower loop ranged at [0.218\emph{J}, 0.232\emph{J}] and the upper loop ranged at [0.232\emph{J}, 0.245\emph{J}]. From the evolution of magnetization texture, one can see that the lower loop forms when parts of the three anti-aligned domains become merged. As $B_{\textup{ext}}$ increases to the value of 0.232\emph{J}, the rest of the three anti-aligned domains start to merge, giving rising to the upper loop. When all of the anti-aligned domains become connected, the three aligned domains appear. Different from the vortex domain wall structure of the initial magnetization texture, the ones in the aligned domains are surrounded by one vortex domain wall with winding number of 1 centered at AA stacking and two anti-vortex domain walls with winding number of $-2$ centered at AB and BA stackings. The main reason of this arrangement of winding number is that, in order to minimize the intralayer exchange energy, the in-plane magnetizations tend to align parallel. With the increase of external magnetic field, the three anti-aligned domains come close and finally merge into each other. In order to keep parallelly aligned at the merging areas, the in-plane magnetizations form a vortex structure with winding number of 1 around the AA stacking domain. While, around the AB and BA domains, the winding numbers become -2. The in-plane magnetizations around the domain wall of AA stacking domain would also tend to align parallel with the nearby in-plane magnetizations around the domain walls of AB and BA stacking domains. This leads to a $\pi/2$ helicity of the in-plane magnetizations around the domain wall of AA stacking domain, which is different from the zero helicity of the initial configuration. For initial states with 1SK and 2SK shown in Figs. 3(d) and (e), we have checked that the winding numbers around the AA AB, and BA domains are \{0,-1,0\} and \{0,-1,-1\} respectively at the end of the magnetic dynamics. The change of anti-aligned to aligned domains, together with the change of domain wall structures, ensures that the topological charge is preserved during the evolution. When decreasing the magnetic field, the magnetization texture returns to its initial configuration. The whole process then realizes a topologically protected magnetic switching. We note that if further increasing the magnetic field to some critical value, where a topological phase transition happens, the topological charge in the whole process would not be conserved anymore.

For CrBr$_3$, which has much weaker anisotropic energy (see Table II) and smaller magnitude of interlayer coupling (see Fig.2 (b)) than CrI$_3$, the magnetization domains are easier to break their original shapes, which results in a much smooth hysteresis loop, as shown in Fig. 10 in Appendix C. Because the aligned domains centered at the AB and BA stackings are formed almost simultaneously, there is only one hysteresis loop in this case. Furthermore, the hysteresis loop is easier to observe in a long-period moir\'{e} pattern with small twisting angle. For a short-period moir\'{e} pattern with large twisting angle, the domain size is small and it costs less intralayer exchange energy for it to merge into each other. Hence, the magnetic dynamics in a short-period moir\'{e} pattern is almost reversible, as shown in Fig. 11 in Appendix C.

\begin{figure}[h]
\centering
\includegraphics[width = 1.0\columnwidth] {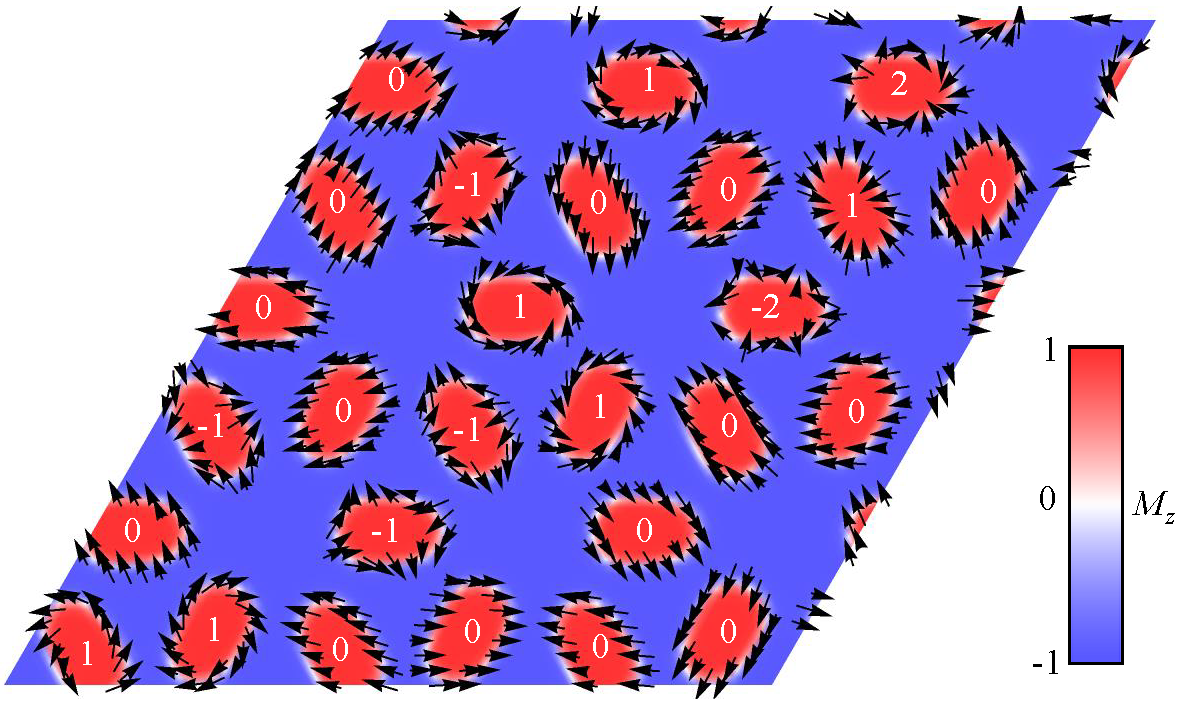}
\caption{ Steady-state magnetization textures evolved from a paramagnetic state for a  $3\times 3$ moir\'{e} supercell of CrI$_3$. The local topological charge is indicated at each domain. The parameter used is  $A=30a$.  } \label{fig6}
\end{figure}

\section{discussions and summary}\label{cd}
The ground state magnetization texture in a long-period moir\'{e} magnet is MB state, which can be spontaneously prepared by first polarizing the magnetization in a spin polarized state with a strong external magnetic field and then ramping down the field for the magnetization to relax to the ground state. To prepare the SK texture, which is excited state, we first prepare the magnetic state to the paramagnetic state, in which the magnetization aligns randomly and then reduce the temperature to a ferromagnetic state. From this annealing process, one can prepare various magnetization configurations. Fig. 6 shows the simulation results evolving from a random magnetization configuration. Besides the SK with vortex texture of $C=1$, there also exists SK with anti-vortex texture of $C=-1$, and magnetization textures of high topological number of $C=\pm 2$ in each local domain. All of these textures are metastable configurations with energies higher than the MB state. Because the initial paramagnetic state, in which all of the magnetic moments align randomly, has no periodicity, the local topological charges after annealing in the steady state have no local periodicity as well. In a long-period moir\'{e} pattern, the local magnetic domains are separate in moir\'{e} space and almost independent. Therefore, it is possible to have neighboring magnetization domains with aperiodic topological index. They would animate into topological trivial MB states when domains with aperiodic topological index come close to each other, for example by applying an external magnetic field.

In summary, we have studied the magnetization textures in the moir\'{e} pattern of twisted bilayer CrX$_3$ (X=Br, I). We show that periodic magnetization domains can be formed arising from the lateral modulation of interlayer magnetic interaction. Magnetization textures with various topological numbers can be generated via annealing from a paramagnetic state. An external magnetic field can be used to tune these magnetic configurations. These intriguing magnetization textures with high controllability not only point to new possibilities to study magnetism in 2D limit, but also provide a potential platform to construct scalable spintronic devices.

\begin{figure}[tbp]
\centering
\includegraphics[width = 1.0\columnwidth] {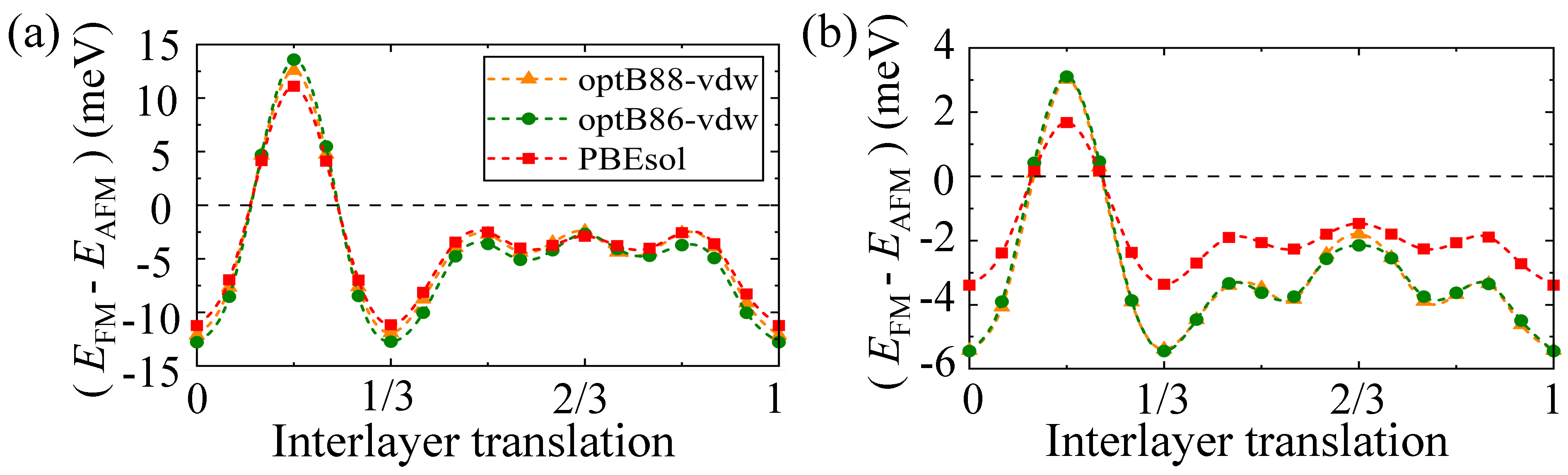}
\caption{ Energy difference between interlayer FM and AFM states of bilayer CrI$_3$ (a) and bilayer CrBr$_3$ (b) with optB88-vdW (orange-dashed line with triangles), optB86-vdW (green-dashed line with dots) and PBEsol (red-dashed line with squares) along the [110] direction. } \label{fig7}
\end{figure}

\section*{ACKOWLEDGEMENTS}
This work is supported by the National Natural Science Foundation of China (Grants No. 11904095) and the Fundamental Research Funds for the Central Universities from China.

\section*{APPENDIX A. THE FIRST-PRINCIPLES CALCULATIONS}
First-principles calculations are implemented in the Vienna Ab initio Simulation Package (VASP) with the PBEsol functional \cite{Kresse1996,Perdew2008}. In order to account for strong electronic correlations for the Cr atoms, a Hubbard on-site Coulomb parameter of 3eV was selected in the calculations \cite{Liechtenstein1995}. A vacuum layer with thickness of $15\textup{\AA }$ is used to eliminate the interaction between the layers. The convergence criteria for energy and force are set as $10^{-5}\textup{eV}/\textup{\AA }$ and $10^{-3}\textup{eV}/\textup{\AA }$, the Monkhorst-Pack mesh is set to $12\times 12\times 1$, and a plane-wave cutoff energy of 450eV was used in the calculations. The calculated structure parameters of AB-stacking bilayer CrX$_3$ (X=Br, I) are showed in the Table I. For simplicity, in studying the translation dependence of interlayer magnetic interaction in Fig. 2, the structure parameters are fixed as the one in AB-stacking. The intralayer exchange coupling \emph{J} and magnetic anisotropic energy \emph{K} of the monolayer CrX$_3$ (X=Br, I) are given in Table II.

\begin{table}[htb]
\centering
\caption{ The structure parameters of AB-stacking CrX$_3$ (X=Br, I) obtained by first-principles calculations. \emph{a} is the lattice constant, \emph{b} is Cr-X intralayer distance, \emph{c} is Cr-Cr intralayer distance, and \emph{d} is interlayer distance between two planes containing the Cr atoms. All units are given in $\textup{\AA }$. }
\begin{tabular*}{\hsize}{@{}@{\extracolsep{\fill}}lllllllllllll@{}}
\toprule
 &\emph{a}&\emph{b}&\emph{c}&\emph{d}\\
\colrule
CrI$_3$&6.83&2.70&3.94&6.61\\
CrBr$_3$&6.30&2.48&3.64&6.27\\
\botrule
\end{tabular*}
\label{symbols}
\end{table}

\begin{table}[htb]
\centering
\caption{ Intralayer exchange coupling $\emph{J}$ and magnetic anisotropic energy $\emph{K}$ in monolayer CrX$_3$ (X=Br, I). The magnetic anisotropic energies are adopted from the experimental results of bulk chromium trihalides. $^{\rm a}$Ref.\cite{Tong2018}, $^{\rm b}$Ref.\cite{Dillon1962} and $^{\rm c}$Ref.\cite{McGuire2015}. }
\begin{tabular*}{\hsize}{@{}@{\extracolsep{\fill}}lllllllllllll@{}}
\toprule
 &$J(\textup{mev})$&$K/J$\\
\colrule
CrI$_3$&1.6$^\textup{a}$ &0.057$^\textup{c}$\\
CrBr$_3$&1.5$^\textup{a}$ &0.016$^\textup{b}$\\
\botrule
\end{tabular*}
\label{symbols}
\end{table}

\begin{figure*}[htb]
\centering
\includegraphics[width = 1.3\columnwidth] {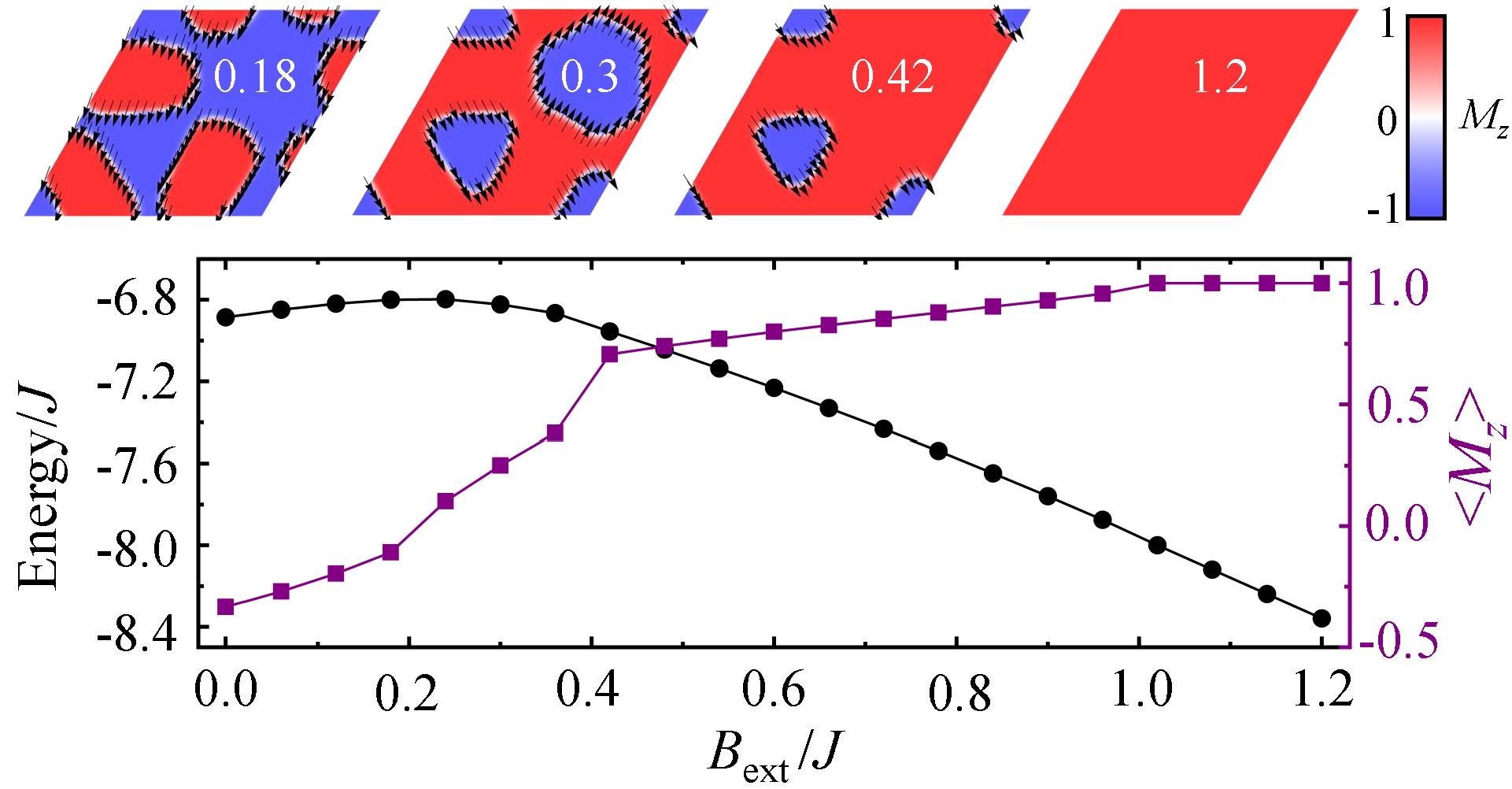}
\caption{ Energy (black line with diamonds) and averaged out-of-plane magnetization  $\left \langle M_{z} \right \rangle$ (purple line with squares) for 3MB magnetization texture of twisted bilayer CrI$_3$ in a moir\'{e} supercell with $A=57a$  under an external magnetic field along \emph{z}-direction. Four magnetization textures of $B_{\textup{ext}}/J=\left \{ 0.18,0.3,0.42,1.02 \right \}$  are shown on the top panel.  } \label{fig8}
\end{figure*}

\begin{figure*}[htb]
\centering
\includegraphics[width = 1.3\columnwidth] {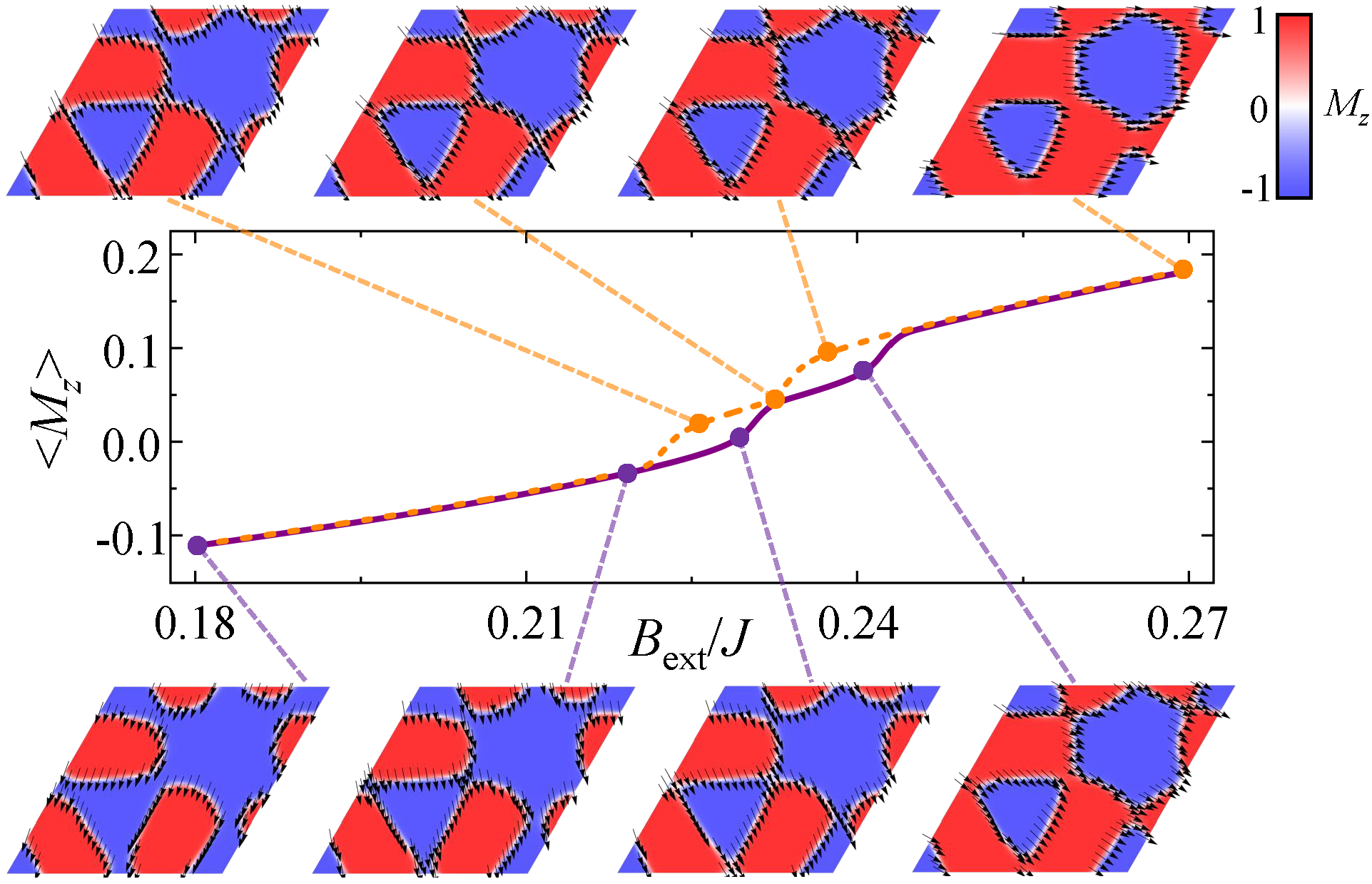}
\caption{ Magnetic hysteresis for 3MB magnetization texture of twisted bilayer CrI$_3$ in a moir\'{e} supercell with $A=57a$  when slowly increasing (purple-solid line) and then decreasing (orange-dashed line) external magnetic field. The magnetic configurations at representative magnetic field values are plotted.   } \label{fig9}
\end{figure*}

\begin{figure*}[htb]
\centering
\includegraphics[width = 1.3\columnwidth] {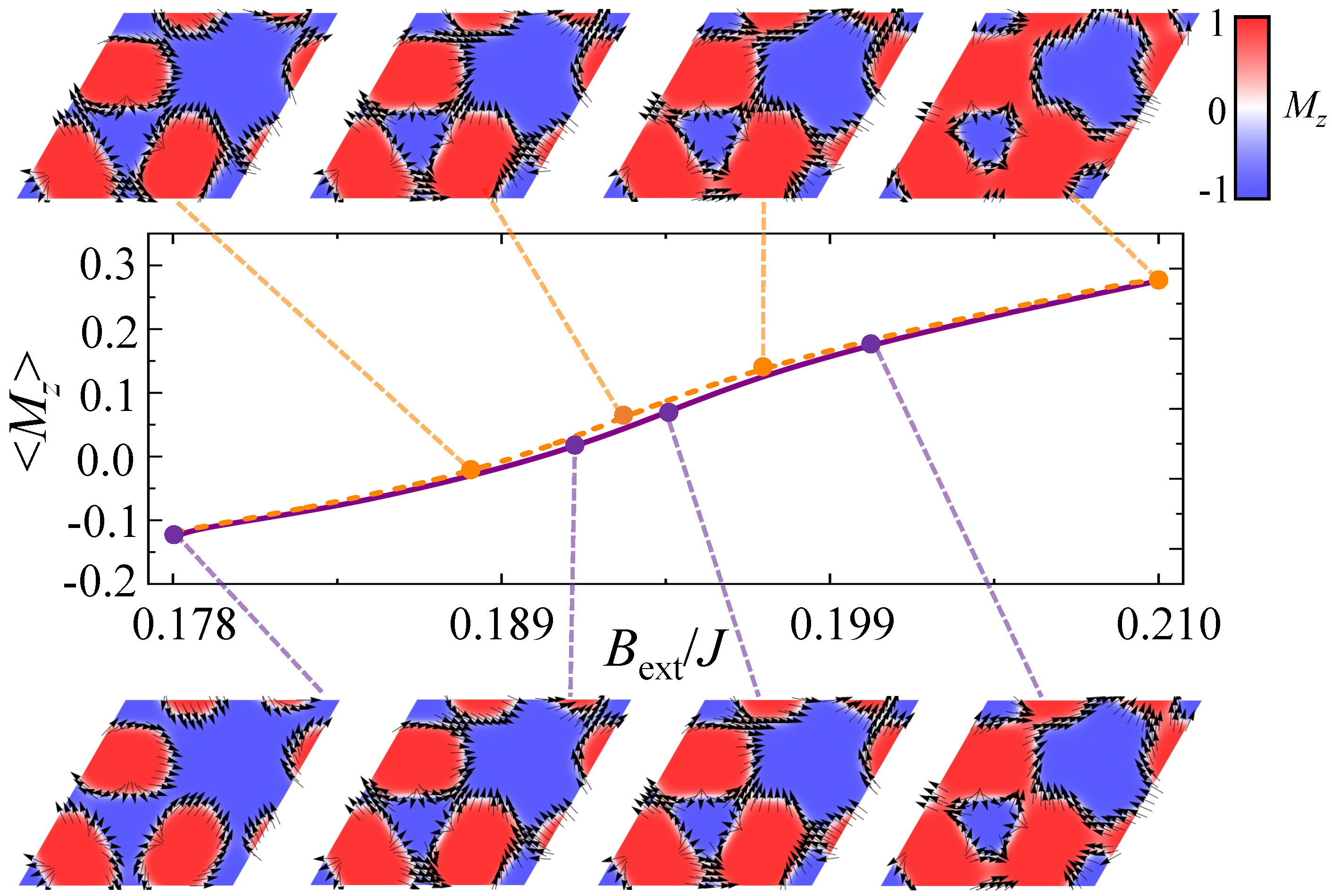}
\caption{ Magnetic hysteresis for 3SK magnetization texture of twisted bilayer CrBr$_3$ in a moir\'{e} supercell with $A=57a$  when slowly increasing (purple-solid line) and then decreasing (orange-dashed line) external magnetic field. The magnetic configurations at representative magnetic field values are plotted.  } \label{fig10}
\end{figure*}

\begin{figure*}[htb]
\centering
\includegraphics[width = 1.3\columnwidth] {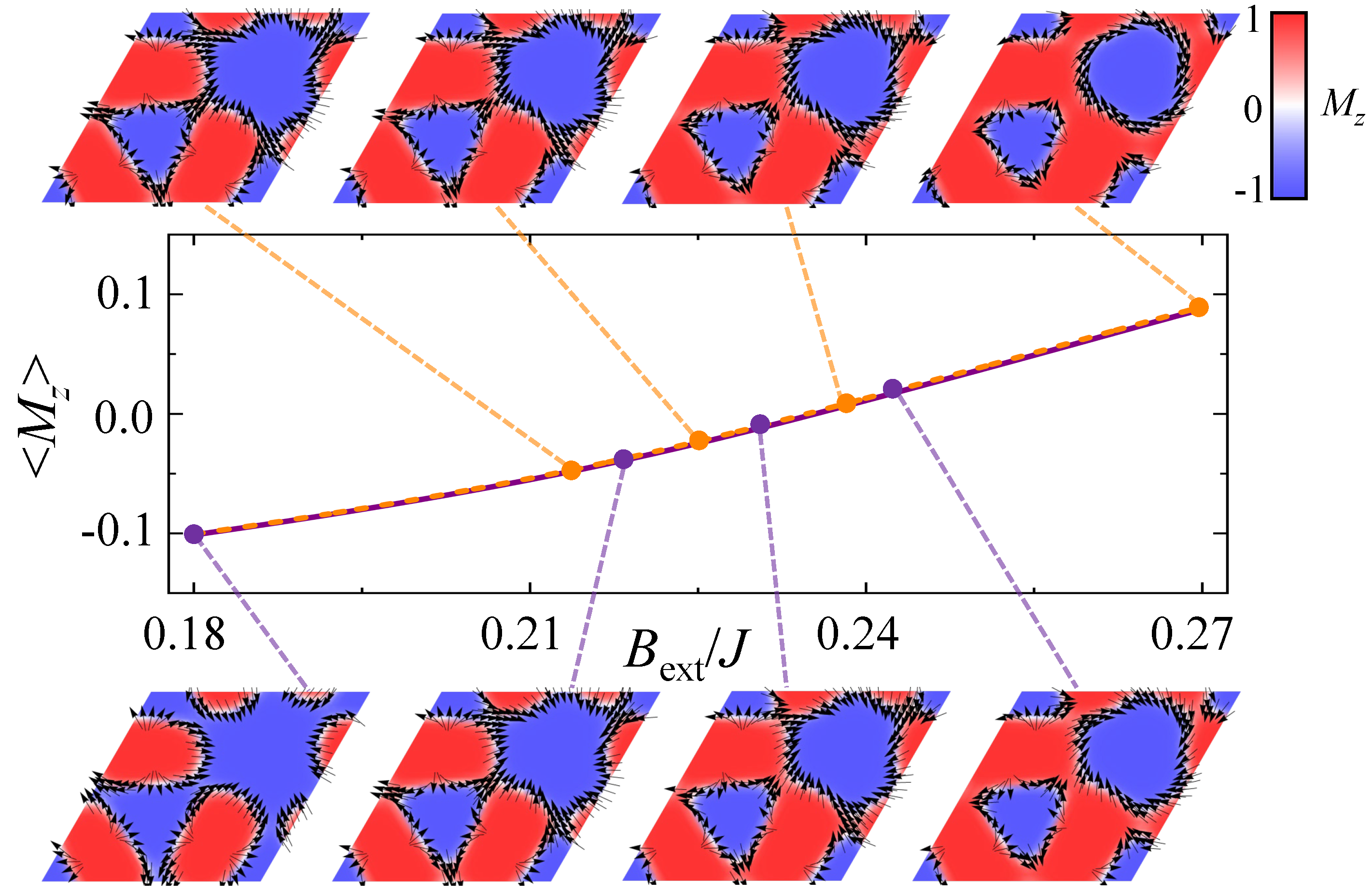}
\caption{ Magnetic hysteresis for 3SK magnetization texture of twisted bilayer CrI$_3$ in a moir\'{e} supercell with $A=30a$  when slowly increasing (purple-solid line) and then decreasing (orange-dashed line) external magnetic field. The magnetic configurations at representative magnetic field values are plotted.  } \label{fig11}
\end{figure*}

To see how the interlayer magnetic interaction depend on the vdW corrections, we perform calculations using optB88-vdW and optB86-vdW functionals \cite{JK2011}. From Fig. 7, one can see that the magnitudes of the energy differences between interlayer FM and AFM states are enhanced with the inclusion of vdW corrections, which would change the moir\'{e} magnetic field quantatively and the shapes of the magnetization textures would reconfigure accordingly. However, the main results presented in this work would not change, i.e. the spatially modulated moir\'{e} magnetic field would create magnetization textures in a long-period moir\'{e} pattern.

\section*{APPENDIX B. MAGNETIC FIELD CONTROL OF 3MB MAGNETIZATION TEXTURES}
In the main text, we have studied the magnetic control of 3SK magnetization texture. We find that other textures show similar magnetic field dependent behavior. In Figs. 8 and 9, we present results for 3MB magnetization texture. Although the topological number is different, the energy, averaged magnetization and its dynamics show similar behavior as the 3SK magnetization texture.

\section*{APPENDIX C. MAGNETIC FIELD CONTROL OF TWISTED BILAYER CRBR$_3$ AND SHORT-PERIOD MOIR\'{E} PATTERN}
In Figs. 10 and 11, we study the magnetic field control of twisted bilayer CrBr$_3$ and short-period moir\'{e} pattern. The much weaker anisotropic energy and smaller magnitude of interlayer coupling in CrBr$_3$ make the magnetization domain easier to break its original shape, resulting in a much smooth hysteresis loop (Fig. 10). For a short-period moir\'{e} pattern, the domain size is small and it costs less intralayer exchange energy to merge into each other. Therefore the magnetic dynamics in a short-period moir\'{e} pattern is almost reversible (Fig. 11).

\end{document}